\documentclass[aps,twocolumn,pra,showpacs,floatfix,superscriptaddress]{revtex4-1}

\usepackage{amssymb}
\usepackage{amsmath}
\usepackage{graphicx}
\usepackage{dcolumn}

\newcommand{\be}{\begin{eqnarray}}
\newcommand{\ee}{\end{eqnarray}}
\newcommand{\la}{\langle}
\newcommand{\ra}{\rangle}

\newcommand{\veps}{\varepsilon}

\newcommand{\rmd}{{\rm d}}

\newcommand{\tm}{\tablenotemark}
\begin{document}

\title{\bf Ground-state hyperfine splitting for Rb, Cs, Fr, Ba$^{+}$, and Ra$^{+}$}

\author{J. S. M. Ginges}
\affiliation{Centre for Engineered Quantum Systems, School of Physics,
The University of Sydney, Sydney NSW 2006, Australia}
\author{A. V. Volotka}
\affiliation{Helmholtz-Institut Jena, Fr\"obelstieg 3, D-07743 Jena, Germany}
\author{S. Fritzsche}
\affiliation{Helmholtz-Institut Jena, Fr\"obelstieg 3, D-07743 Jena, Germany}
\affiliation{Theoretisch-Physikalisches Institut, Friedrich-Schiller-Universit\"{a}t Jena, 
Max-Wien-Platz 1, D-07743 Jena, Germany}

\date{\today}

\begin{abstract}
We have systematically investigated the ground-state hyperfine structure for alkali-metal
atoms $^{87}$Rb, $^{133}$Cs, $^{211}$Fr and alkali-metal-like ions $^{135}$Ba$^{+}$,
$^{225}$Ra$^{+}$, which are of particular interest for parity violation studies. The quantum
electrodynamic one-loop radiative corrections have been rigorously evaluated within an extended
Furry picture employing core-Hartree and Kohn-Sham atomic potentials. Moreover, the effect of
the nuclear magnetization distribution on the hyperfine structure intervals has been 
studied in detail and its uncertainty has been estimated. Finally, the theoretical description of the
hyperfine structure has been completed with full many-body
calculations performed in the all-orders correlation potential method. 

\end{abstract}


\maketitle

\section{Introduction}

It has become apparent in the last decade that quantum electrodynamic (QED) radiative corrections
may give sizeable contributions to properties of heavy atoms. Notably, QED radiative corrections
to the parity violating electric dipole amplitude in Cs were critical in restoring the deviation
of the measured value of the nuclear weak charge from the value predicted by the standard model of
particle physics \cite{wieman,Sush_ueh,JBS_ueh,KF2002,MST2002,SPVC2003,SPTY2005,STPPY2005,rad_pot,review}.
Indeed, it has been shown that QED radiative corrections may contribute at the level (0.1-1)\% to
$s$-$p$ energy intervals \cite{LGTP1999,SC_lamb,rad_pot,TS2010,RDF2013,mod_pot,GB2016}, usual E1
amplitudes \cite{SC_e1,rad_pot,RDF2013,SD2017}, hyperfine structure
intervals \cite{SC_hfs,SC_hfs_p3}, 
and parity violating amplitudes \cite{SPTY2005,STPPY2005,rad_pot,RDF2013,SD2017} for heavy
alkali-metal atoms and near-neutral alkali-metal-like ions. 

The current work is motivated by a new generation of atomic parity violation experiments that are
underway or in preparation for Cs \cite{cs_pnc}, Fr \cite{fr_pnc1,fr_pnc2,fr_pnc3}, Ba$^{+}$ \cite{ba_pnc1,
ba_pnc2}, and Ra$^{+}$ \cite{ra_pnc}. Rb has also been promoted as a candidate for such studies
\cite{rb_pnc}. Single-isotope measurements of (nuclear spin-independent) atomic parity violation are
sensitive to the nuclear weak charge, which is based on a unique
combination of coupling constants,
and complement collider-based studies \cite{particle1,particle2,particle3}.

Clean interpretation of single-isotope measurements depends on the ability to calculate the atomic
parity violating amplitudes to high accuracy. The highest accuracy has
been reached for Cs, with a claimed 
uncertainty within 0.5\% \cite{pnc_ginges,rad_pot,porsev_pnc,dzuba_pnc}. It is possible to sidestep
atomic theory by performing measurements along an isotope chain and taking ratios of measured values
\cite{ratios}. This approach, however, probes a different combination
of coupling constants than probed in single-isotope studies, 
and the latter studies thus remain valuable in providing unique information
about particle physics \cite{fortson_ratios,ramsey-musolf_APV}.

In the current work, we calculate the ground-state hyperfine structure for atoms of interest for
parity violation measurements. The hyperfine structure depends on the electronic wave functions close
to the nucleus, and comparison between theory and experiment allows the quality of the wave functions
in this region to be gauged. This comparison, along with those for electric dipole transition amplitudes and energy
intervals, forms part of the error analysis for parity violation calculations (see, e.g.,
Refs.~\cite{pnc_ginges,porsev_pnc}). 
In order to make a meaningful comparison for the hyperfine structure, however, one first needs to separate out all other effects. In
particular, the QED radiative and the nuclear magnetization distribution effects need to be accounted
for before an assessment of the quality of the many-electron wave
functions may be made. In this paper we address this need.

The only rigorous QED calculations of the ground-state hyperfine structure for heavy alkali-metal atoms
have been performed by Sapirstein and Cheng \cite{SC_hfs}, and there are no such data for the 
alkali-metal-like ions. Sapirstein and Cheng used the Kohn-Sham
approximation to model the atomic potential. 
In the current work, we perform rigorous QED calculations in two local
atomic potentials, Kohn-Sham and core-Hartree, applied to the ground-state 
hyperfine structure for $^{87}$Rb, $^{133}$Cs, $^{211}$Fr, $^{135}$Ba$^+$, and $^{225}$Ra$^+$.
We also investigate in the current work the nuclear magnetization distribution
effect -- the so-called Bohr-Weisskopf effect --  within different nuclear
models for these systems. 
It is common practice in many-body calculations of the
hyperfine structure for heavy atoms to adopt the model of the
uniformly-magnetized sphere. The validity of this model, however, is
not well-motivated for odd-nucleon nuclei in particular, and different
magnetization models may give significantly different results. 
For example, we observe a very sizeable correction (0.5\%) to the 
hyperfine structure for $^{133}$Cs when using a 
single-particle model for the nuclear magnetization distribution rather than the sphere. 
To complete the theoretical study of the hyperfine structure, 
we perform state-of-the-art many-body
calculations in the all-orders correlation potential
method.

This paper is organized as follows.
In Section~\ref{sec:hfs}, we determine the zeroth-order hyperfine intervals in core-Hartree and Kohn-Sham
potentials.
Our calculations of QED radiative corrections are presented in Section~\ref{sec:qed}.
We discuss different magnetization models and calculate their effects in Section~\ref{sec:bw}. 
In Section~\ref{sec:mb}, we present results of our many-body
calculations, and give a complete theoretical 
description of the hyperfine intervals.
The results are discussed in Section~\ref{sec:disc} and concluding remarks given in Section~\ref{sec:concl}. 

\section{The hyperfine interval}
\label{sec:hfs}

The magnetic interaction of an atomic electron with the magnetic
dipole moment of the nucleus is given by 
\begin{equation}
\label{eq:hfs}
h_{\rm hfs}= |e|{\boldsymbol \alpha}\cdot {\bf A}({\bf r}) =
\frac{|e|}{4\pi}\frac{{\boldsymbol \mu}\cdot ({\bf r}\times
  {\boldsymbol \alpha})}{r^3}F(r) \ ,
\end{equation}
where ${\boldsymbol \alpha}$ is a Dirac matrix,  
${\bf A}$ is the vector potential of the nucleus, ${\boldsymbol
  \mu}=\mu {\bf I}/I$ is the nuclear
magnetic moment, and ${\bf I}$ is the nuclear spin. We use relativistic units 
$c=\hbar=m=1$, $e^2/(4\pi)=\alpha$ 
throughout unless otherwise stated.
The factor $F(r)$ models the magnetization distribution, and for 
a point nucleus $F(r)=1$. 

We consider the hyperfine splitting in the ground states 
of alkali-metal atoms and alkali-metal-like ions. 
The hyperfine interaction Eq.~(\ref{eq:hfs}) splits the state $^{2}S_{1/2}$
into two levels $I\pm 1/2$. The interval between the levels in the
zeroth-order approximation --  with local atomic potential and 
point-nucleus magnetization -- is given by
\begin{equation}
\nu^{(0)} =
\frac{2}{3}\frac{\alpha^2}{m_p}g_I(2I+1)\int_{0}^{\infty}dr\,G_a(r)F_a(r)/r^2\ .
\end{equation}
Here, $G_a(r)$ and $F_a(r)$ are the upper and lower 
radial components of the Dirac single-electron wave function $\varphi_a$
that satisfies the Dirac equation in the extended Furry
representation,  
\begin{eqnarray}
\label{eq:dirac}
\Big({\boldsymbol \alpha}\cdot {\bf p}+\beta+V_{\rm
  nuc}(r)+V_{\rm scr}(r)\Big) \varphi_a =\epsilon_a \varphi_a \ ;
\end{eqnarray} 
$\beta$ is a Dirac matrix, 
$V_{\rm nuc}(r)$ is the nuclear potential, and $V_{\rm scr}(r)$ is the
local screening (electronic) potential that partially
accounts for the interaction between the valence electron 
and the closed core electrons. For $V_{\rm scr}$, in calculations of 
the QED corrections, we use the core-Hartree (CH) potential and the 
Kohn-Sham (KS) potential derived within density-functional theory \cite{KS}. 

In the core-Hartree approach, the $N-1$ core electrons 
are solved self-consistently in the direct potential formed from the
core, 
\begin{equation}
\label{eq:CH}
V_{\rm CH}(r)= \alpha \int_0^{\infty} dr'\, \frac{\rho_{\rm c}(r')}{r_{>}}\ ,
\end{equation}
where $r_{>}=\max(r,r')$ and $\rho_{\rm c}(r)=\sum_b(G^2_b(r)+F_b^2(r))$ is the 
charge density of the core electrons $b$,
$\int_0^\infty dr\rho_{\rm c}(r)=N-1$. 
The valence electron wave functions are found in this potential. In the Kohn-Sham approach \cite{KS}, an 
approximation for the exchange potential is included, 
\begin{equation}
\label{eq:KS}
V_{\rm KS}(r)=\alpha\int_0^{\infty} dr'\, \frac{\rho_{\rm
    t}(r')}{r_{>}}  - \frac{2}{3}\frac{\alpha}{r}\Big[\frac{81}{32\pi^2}\, 
r\rho_{\rm  t}(r)\Big]^{1/3} ,
\end{equation}
where $\rho_{\rm t}$ is the total (core and valence) electron 
density $\rho_{\rm t}=\rho_{\rm c}+(G_a^2(r)+F_a^2(r))$, 
and the self-consistency procedure is carried out in the potential 
formed from all electrons. In the KS approach,
the correct asymptotic form of the atomic
potential at large distances, $V_{\rm nuc}+V_{\rm KS}=-\alpha/r$, is enforced using the Latter correction \cite{latter}.

\begin{table}[bth]
\caption{Nuclear parameters used in this work: root-mean-square radii
  $r_{\rm rms}$ in units ${\rm fm}$, magnetic moments in units $\mu_N$, and
  spin and parity $I^{\pi}$. }
\label{tab:nuclear}
\begin{ruledtabular}
\begin{tabular}{lccccc}
              & $^{87}$Rb  & $^{133}$Cs & $^{135}$Ba & $^{211}$Fr & $^{225}$Ra \\ \hline
$r_{\rm rms}$ & 4.1989     & 4.8041     & 4.8294     & 5.5882     & 5.7150     \\
$\mu$         & 2.751818(2)& 2.582025(3)& 0.838627(2)& 4.00(8)    &-0.7338(15) \\  
$I^{\pi}$     & ${3/2}^{-}$& ${7/2}^{+}$& ${3/2}^{+}$& ${9/2}^{-}$& ${1/2}^{+}$\\
\end{tabular}
\end{ruledtabular}
\end{table}
 
We use a finite nuclear charge potential $V_{\rm nuc}(r)$ at all stages of our calculations,
with charge density corresponding to the two-parameter Fermi distribution, 
\begin{equation}
\rho_{\rm nuc}(r)=\frac{\rho_0}{1+\exp[(r-c)/a]} \ .
\end{equation}
The thickness parameter $t=a(4\ln3)$ is taken to be $t=2.3\, {\rm fm}$ for all nuclei and
the half-density radius $c$ is found from the root-mean-square 
radius $r_{\rm rms}$ compiled in Ref. \cite{rms}, $c^2\approx (5/3)r_{\rm rms}^2 -  (7/3) (\pi  a)^2$.
The isotopes we consider in this work with associated nuclear radii
$r_{\rm rms}$ and nuclear 
moments $\mu$, spin $I$, and parity $\pi$ -- from
Ref.~\cite{moments} -- 
are presented in Table~\ref{tab:nuclear}.

We parameterize the finite-nucleus magnetization and QED radiative
corrections to the hyperfine structure intervals as 
\begin{equation}
\label{eq:hfs_scaling}
\nu=\nu^{(0)}\Big( 1+\frac{\alpha}{\pi} F^{\rm BW}+\frac{\alpha}{\pi} F^{\rm QED}
\Big) \ .
\end{equation}
The finite-nucleus magnetization correction -- the Bohr-Weisskopf (BW)
effect \cite{BW} -- is expressed in terms of a relative correction $F^{\rm BW}$, and 
$F^{\rm QED}$ is the relative QED radiative
correction comprised of the vacuum polarization and self-energy, 
$F^{\rm QED}=F^{\rm VP}+F^{\rm SE}$. 

\begin{table}[htb]
\caption{Zeroth-order hyperfine structure intervals $\nu^{(0)}$ for the ground states of Rb,
  Cs, Ba$^{+}$, Fr, Ra$^{+}$ in core-Hartree and Kohn-Sham potentials. Units: MHz.}
\label{tab:MHz}
\begin{ruledtabular}
\begin{tabular}{lccc}
              & \multicolumn{3}{c}{$\nu ^{(0)}$}    \\
              &{\rm CH}  & {\rm KS} & {\rm KS}\tm[1]\\ \hline
$^{87}$Rb     &  4956.04 &  4886.78 &  4886.320     \\ 
$^{133}$Cs    &  6156.85 &  6164.90 &  6164.831     \\
$^{135}$Ba$^+$&  5652.21 &  5675.51 &               \\
$^{211}$Fr    & 27023.9 & 27545.4 & 27244.2\tm[2] \\
$^{225}$Ra$^+$&-20590.1 &-21128.2 &               \\
\end{tabular}
\end{ruledtabular}
\tablenotetext[1]{Reference \cite{SC_hfs}.}
\tablenotetext[2]{From Ref.~\cite{SC_hfs} obtained
  with isotope $^{212}$Fr and adjusted for different $\mu$.}
\end{table}

Our zeroth-order results $\nu^{(0)}$ in core-Hartree and Kohn-Sham
approximations alongside the values of Sapirstein and Cheng
\cite{SC_hfs} are presented in Table~\ref{tab:MHz}. 
Our Kohn-Sham results agree 
precisely with those of Ref. \cite{SC_hfs} 
when we take the same nuclear parameters (nuclear charge radii, 
nuclear moments) used in that work.
The spread in core-Hartree and Kohn-Sham values for $\nu^{(0)}$ 
for the considered systems is within $3\%$. 

\section{QED radiative corrections}
\label{sec:qed}

The one-loop QED contributions to the hyperfine splitting 
incorporate the self-energy and vacuum polarization
corrections. While there are a number of {\it ab initio}
QED calculations of these corrections for hydrogen-like, lithium-like, and boron-like
ions (see, e.g., \cite{hci_review}), there are only a few
works devoted to the case of neutral heavy atoms. The latter were performed
by Sapirstein and Cheng in alkali-metal atoms for $s$-states in Ref.~\cite{SC_hfs}, 
for $p_{1/2}$-states in Ref.~\cite{SC_hfs_p1},
and for $p_{3/2}$-states in Ref.~\cite{SC_hfs_p3}.
Calculations of QED corrections in neutral atoms must account
for electron screening effects from the very beginning. Thus, in contrast to highly
charged ions where one can use the pure Coulomb potential as the zeroth-order
approximation (the original Furry picture), for neutral atoms the calculations
begin with a local screening potential (the extended Furry picture). 
In this work we employ the core-Hartree and Kohn-Sham potentials,
Eqs.~(\ref{eq:CH}) and (\ref{eq:KS}), repectively.
In this section we evaluate the self-energy and
vacuum polarization corrections within the extended Furry representation for the
hyperfine structure intervals for the ground states 
of Rb, Cs, Ba$^{+}$, Fr, and Ra$^{+}$.

\begin{figure}[hbt]
\includegraphics[width=1.\columnwidth]{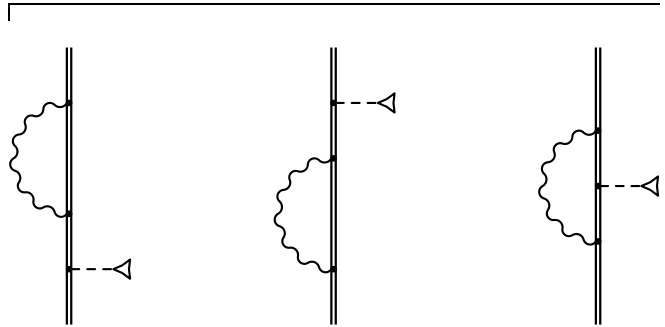}
\caption {Feynman diagrams representing the self-energy correction to the
hyperfine splitting. The wavy line indicates the photon propagator and the
double line indicates the bound-electron wave functions and propagators in the
effective potential comprised of the Coulomb and screening potentials. The
dashed line terminated with the triangle denotes the hyperfine interaction.}
\label{fig:se}
\end{figure}

\begin{figure}[hbt]
\includegraphics[width=1.\columnwidth]{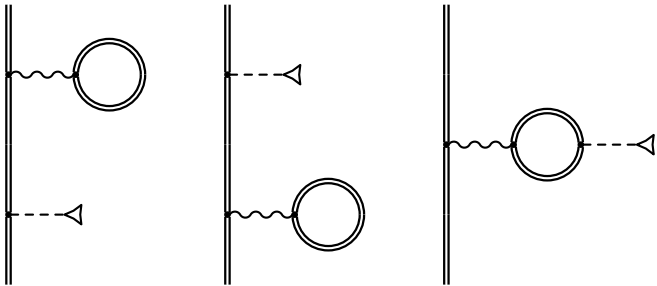}
\caption {Feynman diagrams representing the vacuum polarization correction to
the hyperfine splitting. Notations are the same as in Fig.~\ref{fig:se}.}
\label{fig:vp}
\end{figure}

The complete gauge invariant set of diagrams that need to be considered are
shown in Fig.~\ref{fig:se} and Fig.~\ref{fig:vp} for the self-energy and
vacuum polarization corrections, respectively.
The formal expressions for these diagrams from the first principles of
QED are derived by employing the two-time Green's function method \cite{shabaev_review}.
The correction due to the self-energy diagrams may be written as
\begin{widetext}
\begin{equation}
\label{se}
\nu^{\rm SE} = 
    2 \sum_n^{\veps_n\neq\veps_a} \frac{\la a|T_0|n\ra
     \la n | \Sigma(\veps_a) | a \ra}{\veps_a-\veps_n}
 +  \la a|\frac{\rmd \Sigma(\veps)}{\rmd\veps}\Bigl|_{\veps=\veps_a}|a\ra
     \la a|T_0|a\ra
+ \frac{i}{2\pi} \int^\infty_{-\infty}\rmd \omega \sum_{n_{1} \, n_2}
     \frac{\la a n_2|I(\omega)|n_1 a\ra \, \la n_1|T_0|n_2\ra}
          {(\veps_a-\omega-\veps_{n_1}u)(\veps_a-\omega-\veps_{n_2}u)}\,,
\end{equation}
\end{widetext}
where the first term is the so-called {\it irreducible} part, the
second is the {\it reducible}, and the third one is the {\it vertex} contribution. The self-energy
operator $\Sigma(\veps)$, its derivative $\rmd\Sigma(\veps)/\rmd\veps$, the
interelectronic-interaction operator $I(\omega)$, and the hyperfine operator
$T_0$ are defined in a similar way as in 
Refs.~\cite{volotka2008,volotka2009,glazov2010}, and 
$u=1-i0$ preserves the proper treatment of poles of the electron propagators.
The self-energy corrections given by Eq.~(\ref{se}) suffer from ultraviolet
divergences. In order to cancel these divergences explicitly we have employed
the standard renormalization scheme, details of which may be found,
e.g., in Ref.~\cite{mohr_review}. The infrared divergences which occur in the reducible
and vertex terms are regularized by introducing a non-zero photon mass
and are canceled analytically.

Now let us turn to the vacuum polarization correction to the hyperfine splitting.
The corresponding diagrams are depicted in Fig.~\ref{fig:vp} and provide the following
contribution
\begin{equation}
\label{vp}
  \nu^{\rm VP} =
    2 \sum_n^{\veps_n\neq\veps_a}
     \frac{\la a|T_0|n\ra \la n | U_{\rm VP}^{\rm el} | a \ra}{\veps_a-\veps_n}
 +  \la a | U_{\rm VP}^{\rm ml} | a \ra\,,
\end{equation}
where the first term is the {\it electric-loop} part and the
second is the {\it magnetic-loop} contribution. The electric-field-induced $U_{\rm VP}^{\rm el}$
and the magnetic-field-induced $U_{\rm VP}^{\rm ml}$ vacuum polarization potentials
are defined in a similar way as in Ref.~\cite{andreev2012}. In order to regularize
the ultraviolet divergence terms one has to decompose these potentials into the
Uehling and the Wichmann-Kroll parts. Only the Uehling part
contains the divergent terms, and these may be completely removed using the standard
renormalization procedure \cite{mohr_review}. In this work we have rigorously
evaluated the Uehling parts for both the electric- and magnetic-loop
contributions.
We assume that the ratio of the Uehling and higher-order Wichmann-Kroll terms for
neutral atoms remains similar to the case for hydrogen-like ions. Rigorous
calculations \cite{artemyev2001} of the vacuum polarization correction in hydrogen-like
ions reveal that the Wichmann-Kroll term increases
with nuclear charge and reaches 10\% of the Uehling term for the
heaviest ions considered ($Z=83$). Here, we
do not account for the Wichmann-Kroll terms, since the uncertainty in the treatment of
the screening effects is larger than the estimated contribution of these terms.

\begin{table}[bth]
\caption{Relative QED contributions to the hyperfine splitting of the ground states of
the neutral atoms $^{87}$Rb, $^{133}$Cs, and $^{211}$Fr and the singly-charged ions
$^{135}$Ba$^+$ and $^{225}$Ra$^+$. The vacuum polarization $F^{\rm VP}$, self-energy
$F^{\rm SE}$, and QED $F^{\rm QED} = F^{\rm VP}+F^{\rm SE}$ contributions are presented. 
Our core-Hartree (CH) and Kohn-Sham (KS) results are shown alongside Kohn-Sham results
of Ref.~\cite{SC_hfs}.} 
\label{tab:QED}
\begin{ruledtabular}
\begin{tabular}{lllll}
              &        & $F^{\rm VP}$ & $F^{\rm SE}$ & $F^{\rm QED}$ \\ \hline
$^{ 87}$Rb    &CH      & 0.729        &-1.768        &-1.039         \\
              &KS      & 0.746        &-1.931        &-1.185         \\
              &KS\tm[1]& 0.765        &-1.906        &-1.141         \\
$^{133}$Cs    &CH      & 1.282        &-2.920        &-1.638         \\
              &KS      & 1.323        &-3.229        &-1.906         \\
              &KS\tm[1]& 1.383        &-3.201        &-1.818         \\
$^{135}$Ba$^+$&CH      & 1.305        &-2.906        &-1.601         \\
              &KS      & 1.332        &-3.098        &-1.766         \\
$^{211}$Fr    &CH      & 3.16         &-5.75         &-2.59          \\
              &KS      & 3.31         &-6.43         &-3.12          \\
              &KS\tm[1]& 3.649        &-6.248        &-2.599         \\
$^{225}$Ra$^+$&CH      & 3.15         &-5.50         &-2.35          \\
              &KS      & 3.25         &-5.94         &-2.69          \\
\end{tabular}
\end{ruledtabular}
\tablenotetext[1]{Reference \cite{SC_hfs}.}
\end{table}

In Table~\ref{tab:QED} we present our results for the QED radiative corrections to
the hyperfine structure intervals. Our calculations were performed for finite nuclear
charge (Fermi distribution) and finite nuclear magnetization
(uniformly-magnetized sphere). 
Overall our Kohn-Sham results for Rb, Cs, Fr are in good agreement with the results of
Ref.~\cite{SC_hfs}. For the vacuum polarization correction, the small deviation is due to the finite
nuclear magnetization effect accounted for in our values. For the case of the self-energy, the
calculations are much more involved and the difference can be explained by numerical
uncertainties.
The QED corrections amount to $-0.2\%$ for Rb, $-0.4\%$ for Cs and
Ba$^+$, and $-0.6\%$ for Fr and Ra$^+$. The size of these corrections is significant
and on the level of correlation uncertainties, as we will see in Section~\ref{sec:mb}. 
The variation in results found in CH and KS potentials is within 20\%,
giving an indication of the sensitivity of QED effects to
different treatment of electron screening.

In order to determine the total QED radiative corrections to the
hyperfine splitting, we will apply the same relative QED corrections 
$F^{\rm QED}$ found in the core-Hartree approximation to the final
many-body results presented in Section~\ref{sec:mb}.
We estimate the error associated with this scaling procedure using two methods as follows.
First, based on the results of rigorous calculations of the screened QED radiative corrections for lithium-like
ions \cite{volotka2009,glazov2010} compared to results rescaled from
core-Hartree values, we conservatively estimate the uncertainty as
50\% of the difference between the core-Hartree and rescaled (many-body)
results in absolute units. A second estimate of the uncertainty may be
given as the difference between core-Hartree and Kohn-Sham results. 
The uncertainty we assign to the final QED values is 
the maximum of these estimates.

\section{Bohr-Weisskopf correction}
\label{sec:bw}

For the magnetization distribution, we employ three different models. The first one
is the uniformly magnetized sphere model (sph), where the factor $F(r)$ is given by
\begin{equation}
\label{eq:sphere}
F(r) = (r/r_n)^3 
\end{equation}
for $r \le r_n$ and $F(r) = 1$ elsewhere, where $r_n=(5/3)^{1/2}\,r_{\rm rms}$ is the
nuclear radius. 
The other two are the nuclear single-particle models which are widely used
for the evaluation of the Bohr-Weisskopf correction \cite{lebellac,shabaev,STKAY}.
Within these models the nuclear magnetization is determined by the total angular momentum
of the unpaired proton or neutron. In the first version of this model (SP) we
neglect the nucleon spin-orbit interaction and use a homogeneous distribution for the
radial part of the nucleon wave function inside the nucleus. In the
second version of this model the nucleon spin-orbit interaction is 
included and the nucleon wave function is found in the Woods-Saxon
potential, in a similar way to Ref.~\cite{STKAY}. We consider the latter
model (SP-WS) to be the most reliable, and we will take the results of
this model as our final values for the Bohr-Weisskopf correction. 
We estimate the uncertainty of our SP-WS results as follows. When the contributions
of the nucleon spin and orbital parts are of the same sign, resulting
in a relatively large value for $F^{\rm BW, SP-WS}$, the uncertainty is estimated as
30\% of this value. When the Bohr-Weisskopf correction, evaluated within the model of uniformly magnetized sphere
$F^{\rm BW, sph}$, is larger than that found within the SP-WS model,
the uncertainty is assumed to be 20\% of
$F^{\rm BW, sph}$. To evaluate the Bohr-Weisskopf effect, we use a dense radial grid
with radius $110\,a_B$ ($a_B$ is the Bohr radius) and $10^5$ grid points. The results for the Bohr-Weisskopf
corrections in terms of the $F^{\rm BW}$ factor defined by Eq.~(\ref{eq:hfs_scaling}) are
presented in Table~\ref{tab:BW}. Here, we have used the core-Hartree potential.

\begin{table}[bth]
\caption{Relative BW contributions $F^{\rm BW}$ to the hyperfine splitting of the ground
states of the neutral atoms $^{87}$Rb, $^{133}$Cs, and $^{211}$Fr and the singly-charged
ions $^{135}$Ba$^+$ and $^{225}$Ra$^+$ calculated in the core-Hartree potential.
Results were obtained in three different models of the magnetization
distribution, the uniformly magnetized sphere (sph) and single-particle nuclear models (SP and SP-WS).} 
\label{tab:BW}
\begin{ruledtabular}
\begin{tabular}{lccc}
              & $F^{\rm BW, sph}$ & $F^{\rm BW, SP}$ & $F^{\rm BW, SP-WS}$ \\ \hline
$^{ 87}$Rb    & -1.31     & -1.20    & -1.23(26)   \\
$^{133}$Cs    & -3.07     & -0.89    & -0.80(61)   \\
$^{135}$Ba$^+$& -3.2      & -4.4     & -5.4(16)    \\
$^{211}$Fr    &-11.6      & -5.6     & -6.1(23)    \\
$^{225}$Ra$^+$&-12.1      &-12.1     &-18.7(56)    \\
\end{tabular}
\end{ruledtabular}
\end{table}

It is seen from Tables~\ref{tab:QED} and \ref{tab:BW} that the finite-magnetization correction
tends to make a larger contribution to the hyperfine intervals than the QED radiative
corrections, and this becomes more pronounced for the heavier atoms and ions. Indeed,
for $^{87}$Rb the Bohr-Weisskopf and QED corrections are of comparable size, while for
$^{211}$Fr and $^{225}$Ra$^+$ the Bohr-Weisskopf corrections reach several times the size
of the QED corrections. Moreover, while the sphere model (\ref{eq:sphere}) is frequently used in
calculations of the hyperfine structure for heavy atoms, it is not the most well-motivated
model for odd-nucleon nuclei. 
Indeed, hyperfine structure measurements along a chain of
neutron-deficient isotopes of Fr reveal odd-even staggered results
consistent with a simple nuclear single-particle model
\cite{Fr_chain}. 
For $^{211}$Fr, the hyperfine structure result changes by 1.3\% when
moving from the sphere to the SP-WS model and for $^{133}$Cs it
changes by 0.5\%, a very significant difference on the scale of the
error of the correlation calculations, as we will see in Section~\ref{sec:mb}.

Let us consider the scaling of the relative Bohr-Weisskopf correction in different atomic
potentials. For state $a$, the relative correction is 
\begin{equation}
\label{eq:BW}
\frac{\alpha}{\pi}F^{\rm BW}=
\frac{\int_0^{r_n}dr\, G_a(r)F_a(r)[F(r)-1]/r^2}{\int_{0}^{\infty}dr\,G_a(r)F_a(r)/r^2}
\ ,
\end{equation}
where a spherically symmetric magnetization distribution is assumed. The Bohr-Weisskopf
effect originates on the nucleus where the electron wave functions satisfy the Dirac
equation in the nuclear Coulomb field. For loosely bound valence
electrons with binding energies 
$\ll mc^2, |V_{\rm nuc}|$,  such as we consider in this work, the energy-dependence
is removed and the relativistic radial wave functions for a given angular momentum
quantum number $\kappa$ are the same up to a factor. If the hyperfine interaction were
localized within the $1s$ orbit, we would expect the relative Bohr-Weisskopf correction
(\ref{eq:BW}) for loosely-bound states in different atomic potentials to be very nearly
equal. We have evaluated the Bohr-Weisskopf corrections within both the
core-Hartree and Kohn-Sham potentials 
and indeed observe an approximate equivalence,
\begin{equation}
F^{\rm BW}_{\rm KS}\approx F_{\rm CH}^{\rm BW} \ , 
\end{equation}
to within 0.2\% for the considered atoms and ions.
Furthermore, we have determined the Bohr-Weisskopf correction in both
spherical and  single-particle SP magnetization models at all levels of
many-body approximation (see following section) for $^{133}$Cs and found that $F^{\rm
BW}$ remains the same to within several 0.1\%.
Thus, to find the total Bohr-Weisskopf corrections to the hyperfine splitting,
we will apply the relative values $F^{\rm BW, SP-WS}$ to the final many-body results
presented in the next section.

\section{Many-body calculations}
\label{sec:mb}

\begin{table*}[tbh]
\caption{Results of many-body calculations for the hyperfine structure intervals $\nu$ 
for $^{87}$Rb, $^{133}$Cs, $^{135}$Ba$^+$, $^{211}$Fr, and
$^{225}$Ra$^+$. 
(The hyperfine interval is related to the magnetic constant $A$ by a
factor $(I+1/2)$.)
Relativistic Hartree-Fock (RHF), RHF with all-orders correlation potential $\Sigma^{(\infty)}$,
and RPA with $\Sigma^{(\infty)}$ results -- all with point-nucleus magnetization -- are
presented in the first rows. The correction from semi-empirically adjusting the correlation
potential, $f_{\rm exp}\Sigma^{(\infty)}$, is given in the following row. 
Contributions from structural radiation and normalization ($\delta \Sigma$) and Breit follow.
``Subtotal'' is the sum of the RPA+$\Sigma^{(\infty)}$ value and the
three contributions that follow in the table. 
QED radiative and Bohr-Weisskopf (BW) corrections found in the single-particle magnetization
distribution model (SP-WS) together with their uncertainties are presented in the following rows.
Our final theoretical results are presented as ``Total''. Measured values of the hyperfine
intervals and the deviation ($\Delta$) of theory from experiment in absolute units and in \% are
shown. For the deviation in \% we give in the round brackets the
uncertainty corresponding to the 
QED and BW values. For the case of $^{211}$Fr, an additional
uncertainty associated with the nuclear magnetic moment is presented in the second round brackets.
The final digit in the uncertainty given in brackets matches the final digit for the
central value.  
Units: MHz.}
\label{tab:mb}
\begin{ruledtabular}
\begin{tabular}{lccccc}
                                   &$^{ 87}$Rb   &$^{133}$Cs   &$^{135}$Ba$^+$&$^{211}$Fr      &$^{225}$Ra$^+$ \\ \hline
RHF                                & 4366.1      & 5734.7      & 5252.3       & 29645          &-21865         \\
RHF+$\Sigma^{(\infty)}$            & 5762.8     & 7904.5    & 6343.5& 39438           &-25703         \\
RPA+$\Sigma^{(\infty)}$            & 6878.6      & 9334.8    & 7424.2  & 45824       &-29660         \\
$(f_{\rm exp}-1)\Sigma^{(\infty)}$ &   45.6      &   -2.6      &-10.2     &   9           &   28         \\
$\delta \Sigma$                    &  -95.6      & -126.5      & -144.3     &  -624     &   612         \\
Breit                                    &   11.2      &   23.8      &17.0       &   166           &   -93         \\
Subtotal                           & 6839.8      & 9229.5      & 7286.8     & 45374         &-29113         \\
BW                                 &  -19.5(42)  &  -17.0(131) &  -91.8(275)  &  -641(244)     &  1267(380)    \\
QED                                &  -16.5(23)  &  -35.1(58)  &  -27.1(30)   &  -273(56)      &   159(23)     \\
Total                              & 6803.8      &  9177.4    &7167.9      & 44460          &-27687        \\
Exp.                               & 6834.7\tm[1]& 9192.6\tm[1]& 7183.3\tm[2] & 43570\tm[3]    &-27731\tm[4]   \\ 
$\Delta$                           &  -30.9      &  -15.2      &  -15.4       &   890          &   44         \\
$\Delta$ (\%)                      &   -0.45(7)  &   -0.17(16) &   -0.21(38)  &     2.0(6)(20) &     -0.2(14)   \\
\end{tabular}
\tablenotetext[1]{Reference \cite{rbcs_exp}.}
\tablenotetext[2]{Reference \cite{ba_exp}.}
\tablenotetext[3]{Reference \cite{fr_exp}.}
\tablenotetext[4]{Reference \cite{ra_exp}.}
\end{ruledtabular}
\end{table*}

The hyperfine structure intervals $\nu^{(0)}$ found in the core-Hartree and
Kohn-Sham atomic potentials are very much smaller than the 
measured intervals. For example, for $^{133}$Cs the zeroth-order hyperfine
interval in core-Hartree and Kohn-Sham is around 6200\,MHz, while
the measured value is roughly 9200\,MHz. 
This difference is mostly accounted for by
many-body effects which we address in the current section.

We perform atomic many-body calculations of the hyperfine structure intervals
using the all-orders correlation potential approach \cite{DFSS}. 
The calculations are carried out for point-nucleus
magnetization, and the effects of accounting for finite magnetization
distribution are separately considered (see previous section).
The many-body approximations and methods we use  
have been described at length before, and we
refer the reader to the review \cite{review} for a more detailed description,
diagrams, expressions, and references.

The calculations begin in
the relativistic Hartree-Fock (RHF) approximation, where the local
electronic potential $V_{\rm scr}(r)$ in the Dirac equation (\ref{eq:dirac}) is replaced with the 
RHF potential, 
\begin{equation}
V_{\rm scr}=V^{\rm dir}_{\rm HF}+V^{\rm exch}_{\rm HF} \ ,
\end{equation}
comprised of direct and
exchange parts and formed from the $N-1$ core
electrons. Expressions for this
potential may be found in Ref. \cite{johnson}.
Our RHF values are presented in the first row of results in Table~\ref{tab:mb}.

The choice of RHF as the starting approximation simplifies the
perturbation theory corrections in the residual Coulomb
interaction, with the first non-zero correlation correction appearing
in the second order.
A second-order non-local ``correlation potential'' $\Sigma^{(2)}({\bf
r}_i,{\bf r}_j,\epsilon)$ may be constructed, defined such that its averaged value is equal to the
second-order correlation correction to the energy, $\delta \epsilon
^{(2)}=\langle \varphi |\Sigma^{(2)}|\varphi\rangle$. This potential 
may be added to the RHF equation to
obtain correlation-corrected (Brueckner) wave functions and
energies. 
We go beyond the second order by dressing the Coulomb lines. 
We do this using the Feynman diagram technique to
include important classes of
diagrams -- electron-electron screening and the hole-particle interaction in 
hole-particle loops -- to all orders in the Coulomb interaction \cite{DFSS1988}. 
In this way we obtain an all-orders correlation potential
$\Sigma^{(\infty)}({\bf r}_i,{\bf r}_j,\epsilon)$ which is added to
the RHF equations (\ref{eq:dirac}) with 
\begin{equation}
V_{\rm scr} = V_{\rm HF}+\Sigma^{(\infty)}
\end{equation}
to give Brueckner orbitals $\varphi_{\rm Br}$ and 
energies $\epsilon _{\rm Br}$. 
Consideration of correlation-corrected orbitals corresponds to 
evaluation of the matrix element 
 $\langle \varphi_{\rm Br} |h_{\rm hfs}|\varphi _{\rm
  Br}\rangle$, and the associated hyperfine intervals are presented
in Table~\ref{tab:mb} as RHF+$\Sigma^{(\infty)}$.

Note that in obtaining $\Sigma^{(\infty)}$,
rigorous calculations are performed for the direct diagrams, 
while for the smaller exchange diagrams, simplified second-order calculations
are carried out. 
These latter calculations involve a sum over intermediate states. 
To discretize the states in this sum,
we introduce a cavity of radius $40\,a_B$ and diagonalize the 
relativistic Hartree-Fock Hamiltonian on a set of 40 splines of order 
$k=9$ \cite{johnson}. Higher-order screening corrections are included by 
introducing multipolarity-dependent electron-electron screening 
factors found from direct-diagram calculations.

The random-phase approximation (RPA) with exchange
(time-dependent Hartree-Fock method) is used to account for
polarization of the atomic core by the hyperfine interaction. 
This leads to an additional term in the hyperfine operator \cite{DFSS}, 
\begin{equation}
h_{\rm hfs}\rightarrow h_{\rm hfs}+\delta V_{\rm hfs} \ .
\end{equation} 
This term corresponds to a modification of the RHF
potential with the hyperfine interaction included in first order in the
self-consistency procedure for the core orbitals, 
$\delta V_{\rm hfs}= \tilde{V}_{\rm HF}-V_{\rm HF}$.  We
may express the energy shifts due to the hyperfine interaction,
including correlations and core polarization, as 
$\langle \varphi_{\rm Br} | h_{\rm hfs}+\delta V_{\rm
  hfs}|\varphi_{\rm Br} \rangle$. 
The corresponding results for the hyperfine intervals are shown in
Table~\ref{tab:mb} as RPA+$\Sigma^{(\infty)}$.

We use a simple semi-empirical means of accounting for
missed higher-order correlation corrections. We introduce a factor before 
the correlation potential,  
\begin{equation}
\Sigma^{(\infty)}\rightarrow f_{\rm exp}\Sigma^{(\infty)} \ ,
\end{equation} 
that is found by reproducing experimental binding energies in correlation 
calculations for the energies. 
This also provides us with a good 
indication of the error associated with our many-body calculations.
These semi-empirical corrections are denoted by $(f_{\rm
  exp}-1)\Sigma^{(\infty)}$ in Table~\ref{tab:mb}. 

There are smaller correlation corrections, the ``structural
radiation'', where the hyperfine operator acts on electrons or holes in the 
internal lines of the correlation potential \cite{DFSS}.
We calculate these in the lowest order. 
As with the exchange part of the correlation potential, 
we use splined wave functions in a cavity to calculate the structural
radiation. 
At the same level (third order perturbation theory), there are
corrections to the hyperfine intervals arising from 
normalization of the many-body wave functions \cite{DFSS}, 
$-\langle \varphi |h_{\rm hfs}+\delta V_{\rm hfs}| \varphi \rangle 
\langle \varphi|\partial \Sigma/\partial \epsilon|\varphi\rangle$. These
two corrections 
are bundled together and denoted by $\delta \Sigma$ in Table~\ref{tab:mb}. 

We account for the Breit interaction -- the magnetic and retardation correction to the 
Coulomb interaction --  in the zero-frequency approximation \cite{johnson},
\begin{equation}
h_{\rm Breit}=-\frac{\alpha}{2r}\Big(\boldsymbol{\alpha}_i\cdot \boldsymbol{\alpha}_j+
\boldsymbol{\alpha}_i\cdot {\bf n}\, \boldsymbol{\alpha}_j\cdot {\bf n}\Big),
\end{equation}
where $r$ is the distance between electrons $i$ and $j$. 
Calculations are performed at the RPA level, and the Breit corrections
to hyperfine intervals are given in Table~\ref{tab:mb}.

The contributions described above and tabulated in
Table~\ref{tab:mb} are summed to give the values ``Subtotal''.
These are our final many-body results, for point-nucleus magnetization
and without radiative corrections. 

The Bohr-Weisskopf and QED radiative
corrections are scaled to the many-body values 
and presented, along with their uncertainties (as set out in 
Sections~\ref{sec:qed} and \ref{sec:bw}), in the following rows.
In the final three rows, we give the 
measured values of the hyperfine intervals and the deviations of 
our theoretical results from measurements, in MHz and percent.

\section{Discussion}
\label{sec:disc}

It is seen from Table~\ref{tab:mb} that there is reasonable agreement between theoretical
and experimental values for all considered elements, with agreement 
within several 0.1\% for $^{87}$Rb, $^{133}$Cs, and $^{135}$Ba$^+$.
It is clear, however, that the 
Bohr-Weisskopf uncertainty -- and the nuclear magnetic moment
uncertainty for $^{211}$Fr -- strongly limits testing of the electron wave
functions, as we discuss below.
 
The QED radiative corrections contribute to the hyperfine structure at a level that
is significant and should be taken into account in high-accuracy calculations.
Indeed, the corrections for $^{133}$Cs and $^{135}$Ba$^+$ are both $-0.38\%$, while the
overall deviation of our theoretical determination of the hyperfine structure is 
0.17\% for Cs and 0.21\% for Ba$^+$ (excluding Bohr-Weisskopf
uncertainties). The QED radiative corrections increase with
nuclear charge and contribute $-0.61\%$ and $-0.57\%$ to the hyperfine intervals
for $^{211}$Fr and $^{225}$Ra$^+$, respectively. 

We note that for reliable determination of the QED radiative
corrections to the hyperfine structure, rigorous calculations are
required. In Ref.~\cite{dinh_hfs}, for example, a ``radiative potential''
was used for estimation of the QED effects for the hyperfine
structure. They obtained results for Ba$^+$, Cs, Fr, and Ra$^+$ that
are more than a factor of two larger than those found in the current 
work and Ref.~\cite{SC_hfs}.
A radiative potential -- see, e.g., Refs.~\cite{rad_pot,mod_pot}
-- may be used reliably
for determination of radiative corrections to binding energies and 
other observables where the largest part of the correction arises 
from perturbations to the wave functions (e.g., E1
amplitudes). However, for the hyperfine structure and other short-distance 
operators, there is no reason that the vertex diagrams should be small. 

We have shown that for heavier nuclei the effect of the finite magnetization
distribution becomes increasingly important. In particular, for $^{211}$Fr
and $^{225}$Ra$^+$, its contribution is several times larger than the
QED radiative corrections (by as much as eight times for $^{225}$Ra$^+$). 
Moreover, the uncertainty due to the lack of knowledge of the
magnetization distribution for heavier nuclei completely masks 
the QED radiative corrections, similar to the case for hydrogen-like
ions \cite{STKAY}.
For $^{133}$Cs and $^{135}$Ba$^+$, the Bohr-Weisskopf uncertainty
limits the test of the electronic wave functions by several 0.1\%.
For $^{225}$Ra$^+$, this uncertainty is estimated to be 1.4\%, 
strongly limiting a high-precision
test of many-electron wave functions in hyperfine structure studies.
Furthermore, for $^{211}$Fr, the 2\% uncertainty in the value for the
nuclear magnetic moment prohibits an accurate test.

Depending on the nuclear spin and parity, we emphasise that the 
Bohr-Weisskopf effect may be very different when different
magnetization models are considered. In particular, we have seen for 
$^{133}$Cs that the single-particle model yields a value that 
is several times smaller than the sphere. The result for the ground
state hyperfine splitting changes by as much as 0.5\% when we move 
from the sphere model (commonly used in hyperfine calculations for
this atom) to the more well-motivated nuclear single-particle model.

\section{Conclusion}
\label{sec:concl}

We have performed rigorous calculations of the one-loop QED radiative corrections to the
hyperfine structure intervals for atoms and ions of interest for parity violation studies. 
These corrections contribute $-0.24\%$ ($^{87}$Rb), $-0.38\%$ ($^{133}$Cs), $-0.38\%$
($^{135}$Ba$^+$), $-0.61\%$ ($^{211}$Fr), and $-0.57\%$ ($^{225}$Ra$^+$) and should be included
in accurate theoretical determinations of the hyperfine structure. 
We have also studied the Bohr-Weisskopf correction employing different nuclear magnetization
distribution models and estimated its uncertainty.
We have found that this uncertainty grows with nuclear charge and
strongly impedes the ability to accurately probe the correlation and
QED effects.
We have completed our hyperfine structure analysis with full many-body calculations performed in the
all-orders correlation potential method. 

This work is a step towards an improved theoretical understanding of
the hyperfine structure for heavy atoms and ions. It demonstrates the need for control of the
nuclear physics uncertainties before accurate tests (at the level of 0.1\%) 
of the electronic wave functions in the nuclear region may be made using
hyperfine interval comparisons. 

\section*{Acknowledgments}

J. Ginges is grateful to V. Dzuba for providing the structural
radiation codes and for useful discussions. 
J. Ginges is grateful to Helmholtz-Institut Jena for 
kind hospitality over a three-month visit and also acknowledges 
support from UNSW Sydney where she was based at the start of this
project. 
This work was supported by the Australian Research Council through the 
Centre of Excellence in Engineered Quantum Systems (EQuS), project number CE110001013.


\begin{thebibliography}{999}

\frenchspacing


\bibitem{wieman}

C. S. Wood, S. C. Bennett, D. Cho, B. P. Masterson, J. L. Roberts,
C. E. Tanner, and C. E. Wieman, Science {\bf 275}, 1759 (1997).

\bibitem{Sush_ueh}

O. P. Sushkov, Phys. Rev. A {\bf 63}, 042504 (2001).

\bibitem{JBS_ueh}

W. R. Johnson, I. Bednyakov, and G. Soff, Phys. Rev. Lett. {\bf 87}, 233001 (2001).

\bibitem{KF2002}

M. Yu. Kuchiev and V. V. Flambaum, Phys. Rev. Lett. {\bf 89}, 283002 (2002).

\bibitem{MST2002}

A. I. Milstein, O. P. Sushkov, and I. S. Terekhov,
Phys. Rev. Lett. {\bf 89}, 283003 (2002).

\bibitem{SPVC2003}

J. Sapirstein, K. Pachucki, A. Veitia, and K. T. Cheng, Phys. Rev. A
{\bf 67}, 052110 (2003).

\bibitem{SPTY2005}

V. M. Shabaev, K. Pachucki, I. I. Tupitsyn, and V. A. Yerokhin, 
Phys. Rev. Lett. {\bf 94}, 213002 (2005).

\bibitem{STPPY2005}

V. M. Shabaev, I. I. Tupitsyn, K. Pachucki, G. Plunien, and V. A. Yerokhin, 
Phys. Rev. A {\bf 72}, 062105 (2005).

\bibitem{rad_pot}

V. V. Flambaum and J. S. M. Ginges, Phys. Rev. A {\bf 72}, 052115 (2005).

\bibitem{review}

J. S. M. Ginges and V. V. Flambaum, Phys. Rep. {\bf 397}, 63 (2004).

\bibitem{LGTP1999}

L. Labzowsky, I. Goidenko, M. Tokman, and P. Pyykk\"{o}, Phys. Rev. A
{\bf 59}, 2707 (1999). 

\bibitem{SC_lamb}
J. Sapirstein and K. T. Cheng, Phys. Rev. A {\bf 66}, 042501 (2002).

\bibitem{mod_pot}

V. M. Shabaev, I. I. Tupitsyn, and V. A. Yerokhin, Phys. Rev. A {\bf 88}, 012513 (2013).

\bibitem{TS2010}

C. Thierfelder and P. Schwerdtfeger, Phys. Rev. A {\bf 82}, 062503 (2010).

\bibitem{RDF2013}

B. M. Roberts, V. A. Dzuba, and V. V. Flambaum, Phys. Rev. A {\bf 87},
054502 (2013).

\bibitem{GB2016}

J. S. M. Ginges and J. C. Berengut, Phys. Rev. A {\bf 93}, 052509 (2016).

\bibitem{SC_e1}
J. Sapirstein and K. T. Cheng, Phys. Rev. A {\bf 71}, 022503 (2005).

\bibitem{SD2017}

B. K. Sahoo and B. P. Das, preprint arxiv:1704.04340 (2017).

\bibitem{SC_hfs}

J. Sapirstein and K. T. Cheng, Phys. Rev. A {\bf 67}, 022512 (2003).

\bibitem{SC_hfs_p3}

J. Sapirstein and K. T. Cheng, Phys. Rev. A {\bf 78}, 022515 (2008).

\bibitem{cs_pnc}

D. Antypas and D. S. Elliott, Phys. Rev. A {\bf 87}, 042505 (2013).

\bibitem{fr_pnc1}

E. Gomez, L. A. Orozco, and G. D. Sprouse, Rep. Prog. Phys. {\bf 69}, 79 (2006).

\bibitem{fr_pnc2}

E. Gomez, S. Aubin, R. Collister, J. A. Behr, G. Gwinner,
L. A. Orozco, M. R. Pearson, M. Tandecki., D. Sheng, and J. Zhang, 
J. Phys. Conf. Series {\bf 387}, 012004 (2012).

\bibitem{fr_pnc3}

M.-A. Bouchiat, Phys. Rev. Lett. {\bf 100}, 123003 (2008).

\bibitem{ba_pnc1}

N. Fortson, Phys. Rev. Lett. {\bf 70}, 2383 (1993).

\bibitem{ba_pnc2}

T. W. Koerber, M. Schacht, W. Nagourney, and E. N. Fortson, 
J. Phys. B {\bf 36}, 637 (2003).

\bibitem{ra_pnc}

M. Nu\~{n}ez Portela, E. A. Dijck, A. Mohanty, H. Bekker, J. E. van
den Berg, G. S. Giri, S. Hoekstra, C. J. G. Onderwater, S. Schlesser, 
R. G. E. Timmermans, O. O. Versolato, L. Willmann, H. W. Wilschut, and
K. Jungmann, Appl. Phys. B {\bf 114}, 173 (2014).

\bibitem{rb_pnc}

V. A. Dzuba, V. V. Flambaum, and B. Roberts, Phys. Rev. A {\bf 86}, 062512 (2012).

\bibitem{particle1}

W. J. Marciano and J. L. Rosner, Phys. Rev. Lett. {\bf 65}, 2963
(1990).

\bibitem{particle2}

V. Cirigliano and M. J. Ramsey-Musolf, Prog. in Part. and
Nucl. Phys. {\bf 71}, 2 (2013).

\bibitem{particle3}

J. Erler and S. Su, Prog. in Part. and Nucl. Phys. {\bf 71}, 119 (2013).

\bibitem{pnc_ginges}

V. A. Dzuba, V. V. Flambaum, and J. S. M. Ginges, Phys. Rev. D 
{\bf 66}, 076013 (2002).

\bibitem{porsev_pnc}

S. G. Porsev, K. Beloy, and A. Derevianko, Phys. Rev. Lett. {\bf 102}, 
181601 (2009).

\bibitem{dzuba_pnc}

V. A. Dzuba, J. C. Berengut, V. V. Flambaum, and B. Roberts,
Phys. Rev. Lett. {\bf 109}, 203003 (2012).

\bibitem{ratios}

V. A. Dzuba, V. V. Flambaum, I. B. Khriplovich, Z. Phys. D {\bf 1}, 243 (1986).

\bibitem{fortson_ratios}

E. N. Fortson, Y. Pang, and L. Wilets, Phys. Rev. Lett. {\bf 65}, 2857 (1990). 

\bibitem{ramsey-musolf_APV}

M. J. Ramsey-Musolf, Phys. Rev. C {\bf 60}, 015501 (1999).

\bibitem{KS}

W. Kohn and L. J. Sham, Phys. Rev. {\bf 140}, A1133 (1965).

\bibitem{latter}

R. Latter, Phys. Rev. {\bf 99}, 510 (1955).

\bibitem{rms}

I. Angeli and K. P. Marinova, At. Data Nucl. Data Tables {\bf 99}, 69 (2013).

\bibitem{moments}

N. J. Stone, At. Data Nucl. Data Tables {\bf 90}, 75 (2005).

\bibitem{BW}

A. Bohr and V. F. Weisskopf, Phys. Rev. {\bf 77}, 94 (1950).

\bibitem{hci_review}

A. V. Volotka, D. A. Glazov, G. Plunien, and V. M. Shabaev,
Ann. Phys. {\bf 525}, 636 (2013).

\bibitem{SC_hfs_p1}

J. Sapirstein and K. T. Cheng, Phys. Rev. A {\bf 74}, 042513 (2006).

\bibitem{shabaev_review}

V. M. Shabaev, Phys. Rep. {\bf 356}, 119 (2002).

\bibitem{volotka2008}

A. V. Volotka, D. A. Glazov, I. I. Tupitsyn, N. S. Oreshkina, G. Plunien, and V. M. Shabaev,
Phys. Rev. A {\bf 78}, 062507 (2008).

\bibitem{volotka2009}

A. V. Volotka, D. A. Glazov, V. M. Shabaev, I. I. Tupitsyn, and G. Plunien,
Phys. Rev. Lett. {\bf 103}, 033005 (2009).

\bibitem{glazov2010}

D. A. Glazov, A. V. Volotka, V. M. Shabaev, I. I. Tupitsyn, and G. Plunien, Phys. Rev. A
{\bf 81}, 062112 (2010).

\bibitem{mohr_review}

P. J. Mohr, G. Plunien, and G. Soff, Phys. Rep. {\bf 293}, 227 (1998).

\bibitem{andreev2012}

O. V. Andreev, D. A. Glazov, A. V. Volotka, V. M. Shabaev, and G. Plunien, Phys. Rev. A
{\bf 85}, 022510 (2012).

\bibitem{artemyev2001}

A. N. Artemyev, V. M. Shabaev, G. Plunien, G. Soff, and
V. A. Yerokhin, Phys. Rev. A {\bf 63}, 062504 (2001).

\bibitem{lebellac}

M. Le Bellac, Nucl. Phys. {\bf 40}, 645 (1963).

\bibitem{shabaev}

V. M. Shabaev, J. Phys. B {\bf 27}, 5825 (1994).

\bibitem{STKAY}

V. M. Shabaev, M. Tomaselli, T. K\"uhl, A. N. Artemyev, and
V. A. Yerokhin, Phys. Rev. A {\bf 56}, 252 (1997).

\bibitem{Fr_chain}

J. Zhang, M. Tandecki, R. Collister, S. Aubin, J. A. Behr, E. Gomez,
G. Gwinner, L. A. Orozco, M. R. Pearson, and G. D. Sprouse, 
Phys. Rev. Lett. {\bf 115}, 042501 (2015).

\bibitem{DFSS}

V. A. Dzuba, V. V. Flambaum, P. G. Silvestrov, O. P. Sushkov, J. Phys. B
{\bf 20}, 1399 (1987).

\bibitem{johnson}

W. R. Johnson, {\it Atomic Structure Theory}, Lectures on
Atomic Physics (Berlin, Springer, 2007).

\bibitem{rbcs_exp}

E. Arimondo, M. Inguscio, and P. Violino, Rev. Mod. Phys. {\bf 49},  31 (1977).

\bibitem{ba_exp}

W. Becker, R. Blatt, and G. Werth, J. Phys. (Paris) {\bf 42}, C8-339 (1981).

\bibitem{fr_exp}

J. E. Sansonetti, J. Phys. Chem. Ref. Data {\bf 36}, 497 (2007).

\bibitem{ra_exp}

U. Dammalapati, K. Jungmann, and L. Willmann, J.Phys. Chem. Ref. Data {\bf 45}, 013101 (2016).

\bibitem{DFSS1988}

V. A. Dzuba, V. V. Flambaum, P. G. Silvestrov, O. P. Sushkov,
Phys. Lett. A {\bf 131}, 461 (1988).

\bibitem{dinh_hfs}

T. H. Dinh, V. A. Dzuba, and V. V. Flambaum, Phys. Rev. A {\bf 80},
044502 (2009).

\end{thebibliography}
\end{document}